\documentclass[conference]{IEEEtran}
\usepackage{cite}
\usepackage{amsmath,amssymb,amsfonts}
\usepackage{algorithmic}
\usepackage{graphicx}
\usepackage{textcomp}
\usepackage{xcolor}
\usepackage{textcmds}
\usepackage{enumerate}
\usepackage[hyphens,spaces,obeyspaces]{url}
\usepackage{hyperref}
\usepackage[multiple]{footmisc}
\usepackage{cleveref}
\usepackage[export]{adjustbox}
\usepackage[caption=false, font=footnotesize]{subfig}
\usepackage{tcolorbox}
\def\BibTeX{{\rm B\kern-.05em{\sc i\kern-.025em b}\kern-.08em
    T\kern-.1667em\lower.7ex\hbox{E}\kern-.125emX}}
    
\usepackage{soul}
\usepackage{tikz}
\usetikzlibrary{calc}

\makeatletter
\newif\if@anonymize

\@anonymizetrue    

\if@anonymize
  \newcommand{\highlight@DoHighlight}{
    \fill [outer sep = -15pt, inner sep = 0pt, color=black]
          ($(begin highlight)+(0,8pt)$) rectangle ($(end highlight)+(0,-3pt)$) ;
  }

  \newcommand{\highlight@BeginHighlight}{
    \coordinate (begin highlight) at (0,0) ;
  }

  \newcommand{\highlight@EndHighlight}{
    \coordinate (end highlight) at (0,0) ;
  }

  \newdimen\highlight@previous
  \newdimen\highlight@current
  \newlength{\item@width}

  \DeclareRobustCommand*\anonymize{%
    \SOUL@setup
    \def\SOUL@preamble{%
      \begin{tikzpicture}[overlay, remember picture]
        \highlight@BeginHighlight
        \highlight@EndHighlight
      \end{tikzpicture}%
    }%
    \def\SOUL@postamble{%
      \begin{tikzpicture}[overlay, remember picture]
        \highlight@EndHighlight
        \highlight@DoHighlight
      \end{tikzpicture}%
    }%
    \def\SOUL@everyhyphen{%
      \discretionary{%
        \SOUL@setkern\SOUL@hyphkern
        \SOUL@sethyphenchar
        \tikz[overlay, remember picture] \highlight@EndHighlight ;%
      }{%
      }{%
        \SOUL@setkern\SOUL@charkern
      }%
    }%
    \def\SOUL@everyexhyphen##1{%
      \SOUL@setkern\SOUL@hyphkern
      \settowidth{\item@width}{##1}%
      \makebox[\item@width]{}%
      \discretionary{%
        \tikz[overlay, remember picture] \highlight@EndHighlight ;%
      }{%
      }{%
        \SOUL@setkern\SOUL@charkern
      }%
    }%
    \def\SOUL@everysyllable{%
      \begin{tikzpicture}[overlay, remember picture]
        \path let \p0 = (begin highlight), \p1 = (0,0) in \pgfextra
          \global\highlight@previous=\y0
          \global\highlight@current =\y1
        \endpgfextra (0,0) ;
        \ifdim\highlight@current < \highlight@previous
          \highlight@DoHighlight
          \highlight@BeginHighlight
        \fi
      \end{tikzpicture}%
      \settowidth{\item@width}{\the\SOUL@syllable}%
      \makebox[\item@width]{}%
      \tikz[overlay, remember picture] \highlight@EndHighlight ;%
    }%
    \SOUL@
  }
\else
  \newcommand{\anonymize}[1]{#1}
\fi
\makeatother

\begin{document}

\title{Tracing Vulnerable Code Lineage\\
}

\author{\IEEEauthorblockN{David Reid}
\IEEEauthorblockA{\textit{Department of EECS} \\
\textit{University of Tennessee}\\
Knoxville, USA \\
dreid6@vols.utk.edu}
\and
\IEEEauthorblockN{Kalvin Eng}
\IEEEauthorblockA{\textit{Department of Computing Science} \\
\textit{University of Alberta}\\
Edmonton, Canada\\
kalvin.eng@ualberta.ca}
\and
\IEEEauthorblockN{Chris Bogart}
\IEEEauthorblockA{\textit{Institute for Software Research} \\
\textit{Carnegie Mellon University}\\
Pittsburgh, USA \\
cbogart@cmu.edu}
\and
\IEEEauthorblockN{Adam Tutko}
\IEEEauthorblockA{\textit{Department of EECS} \\
\textit{University of Tennessee}\\
Knoxville, USA \\
atutko@vols.utk.edu}
}

\maketitle

\begin{abstract}
%
This paper presents results from the MSR 2021 Hackathon. Our team investigates files/projects that contain known security vulnerabilities and how widespread they are throughout repositories in open source software.
These security vulnerabilities can potentially be propagated through code reuse even when the vulnerability is fixed in different versions of the code.
We utilize the World of Code~\cite{ma2019world} infrastructure to discover file-level duplication of code from a nearly complete collection of open source software. This paper describes a method and set of tools to find all open source projects that use known vulnerable files and any previous revisions of those files. 
\end{abstract}


\section{Introduction}
Global coalitions of security-focused individuals and organizations often collaborate to identify and publicize security vulnerabilities and fixes in software. One such example is CVE~\cite{cve}, a canonical list of known software vulnerabilities, curated by an international community of volunteers. As a result, databases are published to alert actors in the affected supply chain, from the software's own maintainers, to the maintainers of other packages that depend on the vulnerable code and end users.  

A notable weakness in this vulnerability tracing pipeline is \emph{code cloning}, the practice of copying functionality from one open source project to another, without creating a traceable dependency which has been been shown to propagate bugs~\cite{Mondal2017}. Without tool-readable formal dependencies between projects employing cloned files and the clone's source, it is unlikely that authors of code containing these copied files will become aware of a critical vulnerability present in a copied file.

Comprehensive searchable open source software archives such as the World of Code (WoC)~\cite{ma2019world} create a new opportunity to capture these invisible copies of vulnerable code. The goal of our hackathon project is to determine how widespread cloned files that contain known vulnerabilities are present throughout repositories. As such, we develop a tool to extract possibly cloned files and provide a proof of concept demonstration of how hidden vulnerabilities can be revealed.

\begin{table*}[ht]
\centering
\caption{Many projects have cloned blobs from a vulnerable project before (\qq{Vulnernable projects}) or after (\qq{Safe projects}) a fix has been applied.  Some (\qq{Unknown projects}) have edited vulnerable files, which may or may not have fixed the known vulnerability.}
\begin{tabular}{l|l|l|l|l|l}
\textbf{Project with CVE} & \textbf{CVE}   & \textbf{Vulnerable Blobs} & \textbf{Vulnerable Projects} & \textbf{Safe Projects} & \textbf{Unknown Projects} \\ \hline
\rule{0pt}{2.5ex}RIOT                  & CVE-2017-8289  & 2                     & 1113                     & 1                  & 950                   \\
QEMU                  & CVE-2018-17962 & 1                     & 3767                     & 21                 & 2361                  \\
LZ4                   & CVE-2019-17543 & 1                     & 36284                    & 12042              & 7443                 
\end{tabular}
\label{tab:counts}
\vspace{-2.5mm}
\vspace{-2.5mm}
\end{table*}

\section{Approach} 
We use a four-step approach to determine how widespread cloned files that contain known vulnerabilities are:
(1) identify vulnerable software releases, either by searching CVE~\cite{cve} or searching for the string \qq{CVE} in software repository commit messages;
(2) identify specific revisions of specific files in those releases, using WoC to identify which files were repaired by the CVE fix;
(3) use WoC to generate a list of all \emph{previous} versions of these files, assuming they are potentially vulnerable, and all \emph{subsequent} versions, assuming they include the fix to the vulnerability;
(4) use WoC to trace these two sets of files across the entirety of open source software, identifying projects that still contain the vulnerability.


Using this approach with additional analysis, we produce three lists of open source projects containing projects which: 

\begin{itemize} 
\item Contain a known vulnerability in the current version of the project. 
\item Contained a known vulnerability in a previous version, but that vulnerability has been fixed in the current version. 
\item Contained a known vulnerability in a previous version, the vulnerable files have been modified in the current version, but it is unknown if the modifications fix the vulnerability. 
\end{itemize}


\section{Algorithm}

First, we start with the SHA-1 hash of the commit that fixes a known vulnerability to lookup all blobs related to the commit in WoC.
Next, we use WoC to recursively find all (potentially vulnerable) ancestors and (potentially fixed) descendants of these blobs, using WoC's blob-to-old-blob and old-blob-to-blob mappings, respectively.


For each vulnerable old blob, 
we use WoC’s blob-to-commit mapping to find the commits of projects that contain the blob. It should be noted that these commits may not be the latest change in that project and thus we are unable to determine if the project still contains the vulnerable blob. Hence, we use WoC’s commit-to-head mapping to obtain the head (newest) commit and all blobs of that commit state in a project to identify three cases: (1) if the bad blob is present in the head, we can conclude that the project is at risk and vulnerable; (2) if any of the \qq{fixed} blobs are in the head commit, we presume that the project has been fixed and is safe; (3) if neither vulnerable or fixed blobs are present in the head commit, the project's status is unknown.



\section{Results}
We choose three cases of vulnerabilities from CVE to demonstrate the feasibility of our methodology. In these cases we identify CVEs in which the commit fixing the vulnerability can be readily identified in projects and may only exist in the original project. In Table~\ref{tab:counts}, we present the vulnerability counts in terms of blobs and projects. Many of the \qq{vulnerable} or \qq{unknown} projects containing the vulnerable code appear to descend from old forks of the main project, but in many cases have not been maintained or used. However, not all are abandoned --- some have been recently forked and have recent comments, suggesting that the known vulnerable code is still being actively used.


\subsection{Case 1: RIOT}
RIOT~\cite{riot} is a real-time multi-threading operating system. In CVE-2017-8289, a stack-based buffer overflow vulnerability in the \texttt{ipv6\_addr\_from\_str} function in 
\texttt{ipv6\_addr\_from\_str.c} was described as being present in versions prior to 2017-04-25 and was subsequently fixed 
in a pull request~\cite{riotfix}. In the fix, two files are changed: \texttt{ipv6\_addr\_from\_str.c} to fix the vulnerability and 
\texttt{tests-ipv6\_addr.c} to test the fix. Using our algorithm to determine if the two fixed files are present, we only find 1 non-fork project that is fixed. In contrast, there are 1,113 projects that still contain a pre-fix revision of one of those two files and 950 projects which contain an unknown version of one of the files leaving its fixed status to be unknown. Notably, the unfixed projects appear to be abandoned forks that implement additional functionality suggesting that caution should be used when using outdated forks that contain additional useful functionality.




\subsection{Case 2: QEMU}
QEMU~\cite{qemu} is a generic and open source machine and userspace emulator and virtualizer. It is subject to a buffer overflow vulnerability as described in CVE-2018-17962. The fix is a simple one line change of a size from \texttt{int} to \texttt{size\_t} in the file \texttt{hw/net/pcnet.c} fixed 
on May 30, 2018. 
Looking at versions of \texttt{pcnet.c} prior to the fixing commit, we find 3,767 projects that contain a vulnerable version of the file, 21 projects that contain the fixed version of the file, and 2,361 projects that potentially contain a fixed or vulnerable file. Even though the vulnerability in \texttt{pcnet.c} was fixed more than 2 years ago, there are still many projects that contain a vulnerable version of the file.  Many of the projects containing the vulnerable code are old forks which do not appear to the maintained or used. However, some have been recently forked and have recent comments, indicating that the known vulnerable code is still being actively used and should be fixed.


\subsection{Case 3: LZ4}
LZ4~\cite{lz4} is a widely-used lossless compression algorithm. The reference implementation of LZ4 is subject to a heap-based buffer overflow in releases prior to 1.9.2 as described in CVE-2019-17543. 
The fix is contained in 1 blob in the file \texttt{lib/lz4.c}, which we are unable to find in 36,284 projects. We find 12,042 projects that contained one of the vulnerable blobs in the past, but now contains one of the known good blobs and is therefore no longer vulnerable. We also find that 7,443 repositories contained the vulnerable blob at one point in time, but the status of the file is unknown as it has since been replaced with a file that is unrelated to the fix. The high count of vulnerable and unknown projects among safe projects suggest that many projects utilizing LZ4 should update their libraries.


\section{Future Work}
In terms of enhancing our method, we can reduce our search space by not assuming that \emph{all} previous revisions of a vulnerable file are vulnerable. By employing an algorithm like \emph{SZZ unleashed}~\cite{szz}, we can determine when a bug is first introduced and rule out prior versions of files in our search space. Furthermore, finer-grained detection of code cloning, at the method rather than file level, is likely to detect more copied vulnerabilities~\cite{Mondal2017}. However, this could be a significant computational challenge as it is difficult to identify the programming language, legitimacy, and encoding of a blob. We also find that searching WoC for clones of not only the vulnerable blobs but also all ancestors of the vulnerable blobs causes performance issues due to the large volume of data. We note that more resources to perform more parallelization and caching of intermediate results could also improve performance.



In terms of suggestions to expand WoC, we note that querying commit messages is still quite time consuming and not very user-intuitive when searching for patterns of strings. We suggest that a more powerful and flexible search mechanism for searching commit log messages should be developed which could help us find commit logs that contain the string \qq{CVE}. 


\section {Conclusion}
In this paper, we present a method to trace vulnerable code lineage in any language throughout open source software, leveraging the World of Code infrastructure. We introduce a tool to implement this method and find evidence of significant reuse of outdated code that contains known vulnerabilities. Several cases are presented to show that many projects reuse code with known vulnerabilities, even though the vulnerabilities have been fixed in other projects suggesting that developers should be more aware when reusing code to avoid using exploitable code.
The code produced and our results are available at \href{https://github.com/woc-hack/hemlock}{github.com/woc-hack/hemlock}.

\bibliographystyle{IEEEtran}
\bibliography{main}

\end{document}